\documentclass[%
 reprint,
superscriptaddress,
 amsmath,amssymb,
 aps,
]{revtex4-2}

\usepackage{graphicx}
\usepackage{dcolumn}
\usepackage{bm}

\usepackage{setspace}
\usepackage{textcomp}
\usepackage{gensymb}
\usepackage[version=3]{mhchem}
\usepackage{multirow, makecell}
\usepackage{graphicx}

\begin{document}

\preprint{APS/123-QED}

\title{Magnetism in EuAlSi and the \ce{Eu_{1-\textit{x}}Sr_{\textit{x}}AlSi} Solid Solution}

\author{Dorota~I.~Walicka}
\affiliation{Department of Quantum Matter Physics, University of Geneva, CH-1211 Geneva, Switzerland}
\affiliation{Department of Chemistry, University of Zurich, CH-8057 Zurich, Switzerland}
\affiliation{Laboratory for Neutron Scattering and Imaging, PSI Center for Neutron and Muon Sciences, 5232 Villigen, PSI, Switzerland}

\author{Olivier~Blacque}
\affiliation{Department of Chemistry, University of Zurich, CH-8057 Zurich, Switzerland}

\author{Karolina Gornicka}
\affiliation{Faculty of Applied Physics and Mathematics, Gdansk University of Technology, 80-233 Gdansk, Poland}
\affiliation{Advanced Materials Center, Gdansk University of Technology, 80-233 Gdansk, Poland}
\affiliation{Materials Science and Technology Division, Oak Ridge National Laboratory, Oak Ridge, Tennessee 37831, United States}

\author{Jonathan~S.~White}
\affiliation{Laboratory for Neutron Scattering and Imaging, PSI Center for Neutron and Muon Sciences, 5232 Villigen, PSI, Switzerland}

\author{Tomasz Klimczuk}
\affiliation{Faculty of Applied Physics and Mathematics, Gdansk University of Technology, 80-233 Gdansk, Poland}
\affiliation{Advanced Materials Center, Gdansk University of Technology, 80-233 Gdansk, Poland}

\author{Fabian~O.~von~Rohr}
\affiliation{Department of Quantum Matter Physics, University of Geneva, CH-1211 Geneva, Switzerland}


\begin{abstract}

The magnetic properties of EuAlSi, a compound comprising a honeycomb lattice of Al/Si atoms and a triangular lattice of Eu atoms, are presented. By means of single-crystal X-ray diffraction we find that EuAlSi crystallizes in a \ce{AlB2}-type structure with space group \textit{P}6/\textit{mmm} and unit cell parameters $a$ = 4.2229(10) \r{A}, and $c$ = 4.5268(12) \r{A}. Our magnetic measurements indicate that EuAlSi is a soft ferromagnetic material with $T_{\rm Curie}$ = 25.8 K. The susceptibility follows the Curie-Weiss law at high temperatures, which allowed us to determine the paramagnetic Curie temperature $\theta_P$ = 36.1 K and an effective magnetic moment $\mu_{eff}$ = 8.0 $\mu_B$/Eu. This value is in agreement with the theoretical value of 7.9 $\mu_B$ for Eu$^{2+}$ free ion. Moreover, we have prepared the \ce{Eu_{1-\textit{x}}Sr_{\textit{x}}AlSi} solid solution, where the atoms in the triangular lattice were systematically exchanged, in order to study the evolution of the collective quantum properties from the ferromagnetic EuAlSi towards the superconducting SrAlSi. Across the \ce{Eu_{1-\textit{x}}Sr_{\textit{x}}AlSi} solid solution the unit cell parameters change linearly, following Vegard's law, suggesting that the changes in physical properties are more likely intrinsic, what makes the system reliable for studying composition dependence of the interplay between the crystal structure and physical properties. As the Sr content increases, i.e. $x$ increases, we note a consistent reduction of $\mu_{\rm eff}$ and $T_{\rm Curie}$. Long-range magnetic order in \ce{Eu_{1-\textit{x}}Sr_{\textit{x}}AlSi} persists up to $x$ = 0.95, whereas superconductivity is only observed for samples with $x$ $>$ 0.95. A possible quantum critical point is suggested to exist in the vicinity of to $x$ $\approx$ 0.96, at which the suppression of ferromagnetic order is concomitant with the emergence of superconductivity.


\end{abstract}

\maketitle

\raggedbottom                              
\maketitle
\sloppy

\section{Introduction}\label{sec:introduction}

The hexagonal \ce{AlB2}-type structure, with its characteristic honeycomb layers, has emerged as a versatile platform for investigating a variety of collective quantum properties. Honeycomb layers as a structural motif are prevalent in several prominent quantum materials, including magic-angle graphene \cite{Cao2018}, \ce{MgB2} \cite{Nagamatsu2001}, and a growing number of ternary superconductors \cite{Evans2009, Walicka2021, Tran2021, walicka2023bagage, Walicka2024}. The unique arrangement of atoms within the \ce{AlB2}-type framework has been linked to the manifestation of diverse quantum phenomena, making these compounds a significant subject of study.

The \ce{AlB2}-type structure, consists also of a triangular Al lattice. This triangular lattice fosters geometrically frustrated systems, including magnets with complex or topologically unique spin textures. Notably, magnetic frustration in this structure allows for the hosting of skyrmions even in centrosymmetric materials. A key example is \ce{Gd2PdSi3}, an \ce{AlB2}-related compound with a triangular Gd lattice, known to exhibit a Bloch-type skyrmion lattice phase at $T_{\rm N}$ = 20 K \cite{Kurumaji2019}.

These \ce{$R$2$T$$X$3}-type compounds (\textit{R} = rare-earth element or U, \textit{T} = transition metal, \textit{X} = Si, Ge, Ga, In) consist of a diverse family of phases displaying different quantum states, including spin-glass behavior, magnetic ordering, the Kondo effect, or heavy fermion behavior \cite{Pan2013}. Among these, Eu-based compounds like \ce{Eu2PdSi3} exhibit competing antiferromagnetic and ferromagnetic correlations below 10 K \cite{Mallik1998}, while \ce{Eu2CuSi3} is a ferromagnet with a Curie temperature of $T_{\rm Curie}$ = 34 K and shows significant negative magnetoresistance up to 100 K \cite{Majumdar1999, cao2010}.

Equiatomic Eu-based compounds with the general formula \ce{Eu$T$$X$} (\textit{T} = transition metal, \textit{X} = p-block element) have been studied, with analyses of their structure and magnetic properties \cite{Pttgen2000}. For the \textit{T} = Ga, compounds such as EuGaSi, EuGaGe, and EuGaSn can be synthesized. EuGaSi crystallizes in the \ce{AlB2}-type structure, while EuGaGe and EuGaSn adopt the \ce{YPtAs}-type structure. EuGaSi and EuGaGe exhibit ferromagnetic behavior, whereas EuGaSn is an antiferromagnet. All three compounds have effective magnetic moments close to that of the Eu$^{2+}$ free ion (7.9 $\mu_B$), corresponding to the $4f^7$ electronic configuration of Eu$^{2+}$ \cite{you2007}.

The existence of the EuAlSi phase crystallizing in the \textit{P}6/\textit{mmm} space group with unit cell parameters $a$ = 4.169 \r{A} and $c$ = 4.510 \r{A} was previously documented \cite{Zarechnyuk1982}. However, to our knowledge, no detailed structural refinement or any physical properties have been reported for this compound. Here, we present the synthesis, the crystal structure, as well as the magnetic and thermal properties of EuAlSi. Furthermore, we have synthesized the \ce{Eu_{1-\textit{x}}Sr_{\textit{x}}AlSi} solid solution of ferromagnetic EuAlSi and superconducting SrAlSi, and analyzed in detail the magnetic properties of this system. 

\section{Methods}\label{sec:exp}

\textbf{Synthesis:} Samples were prepared by arc melting stoichiometric amounts of Eu (99.99 \%  ONYXMET), Sr (99.99 \%  Aldrich), Al (99.9995 \% Acros Organic), and Si (99.95 \%  Aldrich). In the procedure, a piece of Zr was co-heated and used as an oxygen getter. The reactants were placed on a water-cooled copper plate and melted three times in an argon atmosphere using a tungsten tip. After each firing, the sample was flipped over for maximum homogeneity. The weight loss during the process was negligible. All the samples are stable in air.

\textbf{Diffraction:} Powder X-ray diffraction (PXRD) patterns were collected on PANalytical Aeris diffractometer equipped with a PIXcel 1D Medipix detector and a Cu X-ray tube in the 2$\theta$ range from 10--90$\degree$. Single-crystal X-ray diffraction (SXRD) data for EuAlSi were collected at 160.0(1) K on a Rigaku OD Synergy/Hypix diffractometer using a molybdenum X-ray radiation source ($\lambda$= 0.71073 Å) from a dual wavelength X-ray source and an Oxford Instruments Cryojet XL cooler. The selected suitable single crystal was mounted using polybutene oil on a flexible loop fixed on a goniometer head and transferred to the diffractometer. Pre-experiment, data collection, data reduction and analytical absorption correction \cite{Clark1995} were performed with the program suite CrysAlisPro \cite{Crysalispro}. Using Olex2 \cite{Dolomanov2009}, the crystal structure was solved with the SHELXT \cite{Sheldrick2015} small molecule structure solution program and refined with the SHELXL program package \cite{Sheldrick2015a} by full-matrix least-squares minimization on F$^2$. PLATON \cite{Spek2009} was used to check the result of the X-ray diffraction analysis.

\textbf{Physical Properties:} Temperature- and field-dependent magnetization measurements as well as heat capacity measurements were performed using a Quantum Design Physical Property Measuring System (PPMS) Evercool II and a Quantum Design PPMS DynaCool, both machines equipped with a vibrating sample magnetometry (VSM) option and a 9 T magnet. For the magnetic measurements, samples of arbitrary shape and mass range between 5--25 mg were used. Samples for heat capacity measurements were mounted on the measurement platform with Apiezon N grease and measured using the two-$\tau$ time-relaxation method.

\section{Results and discussion}\label{sec:results6}

\subsection{Crystal structure of EuAlSi}\label{sec:EuAlSi}

\begin{figure}
\centering
\includegraphics[width=1\linewidth]{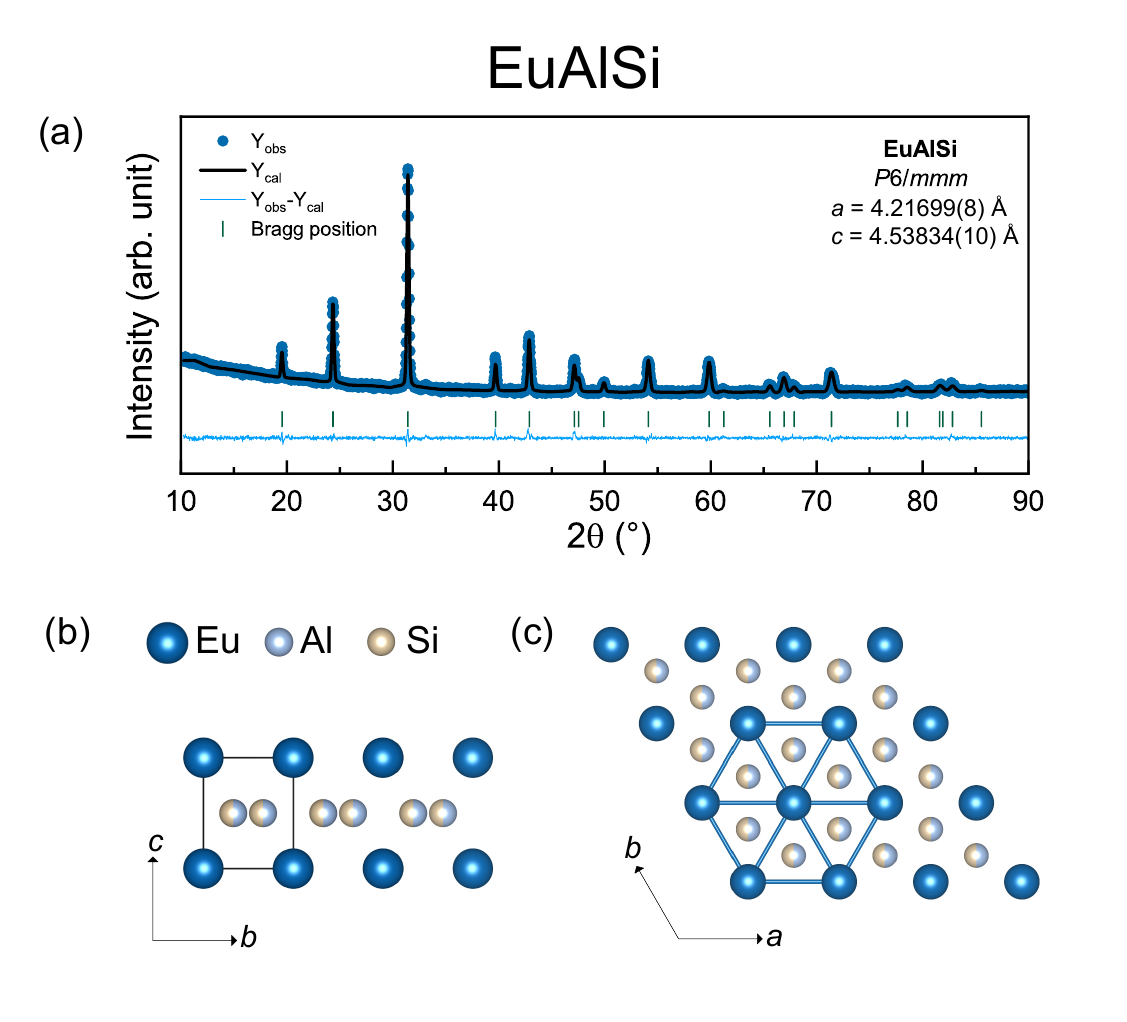}
	\caption{(a) PXRD pattern together with respecting Rietveld refinement for EuAlSi. (b) Crystal structure of EuAlSi along $a$-axis. (c) Crystal structure of EuAlSi along $c$-axis with focus on triangular lattice of Eu.}
	\label{fig:fig1}
\end{figure}

Samples of EuAlSi were obtained as silvery shiny ingots with metallic luster. PXRD data were obtained from finely ground powders of the sample and are shown in Figure \ref{fig:fig1}(a) together with the respecting Le Bail refinement. The perfect agreement of the fit validates the crystal structure of EuAlSi in the space group \textit{P}6/\textit{mmm} with the unit cell parameters $a$ = 4.21699(8) \r{A} and $c$ = 4.53834(10) \r{A}. 

The high crystallinity of the sample enabled us to perform for the first time SXRD measurements on EuAlSi. These experiments confirmed the structural model in the space group \textit{P}6/\textit{mmm}, with unit cell parameters matching the PXRD data: $a$ = 4.2229(10) \r{A} and $c$ = 4.5268(12) \r{A}. The crystal structure of EuAlSi, as determined by SXRD, is depicted in Figure \ref{fig:fig1}(b,c).

The  Al and Si atoms occupy the same crystallographic 2\textit{d} Wyckoff position, forming a honeycomb lattice. The honeycomb layers are sandwiched between layers of Eu atoms, which occupy the crystallographic 1\textit{a} Wyckoff position forming a triangular lattice, with Eu--Eu distance equal to 4.2229(11) \r{A}. The details of the SXRD measurement and the structural refinement are shown in Table \ref{tab:crystal}, while Wyckoff positions, atomic coordinates and occupancy are presented in Table \ref{tab:Wyckoff}.

\begin{figure}
\centering
	\includegraphics[width=1\linewidth]{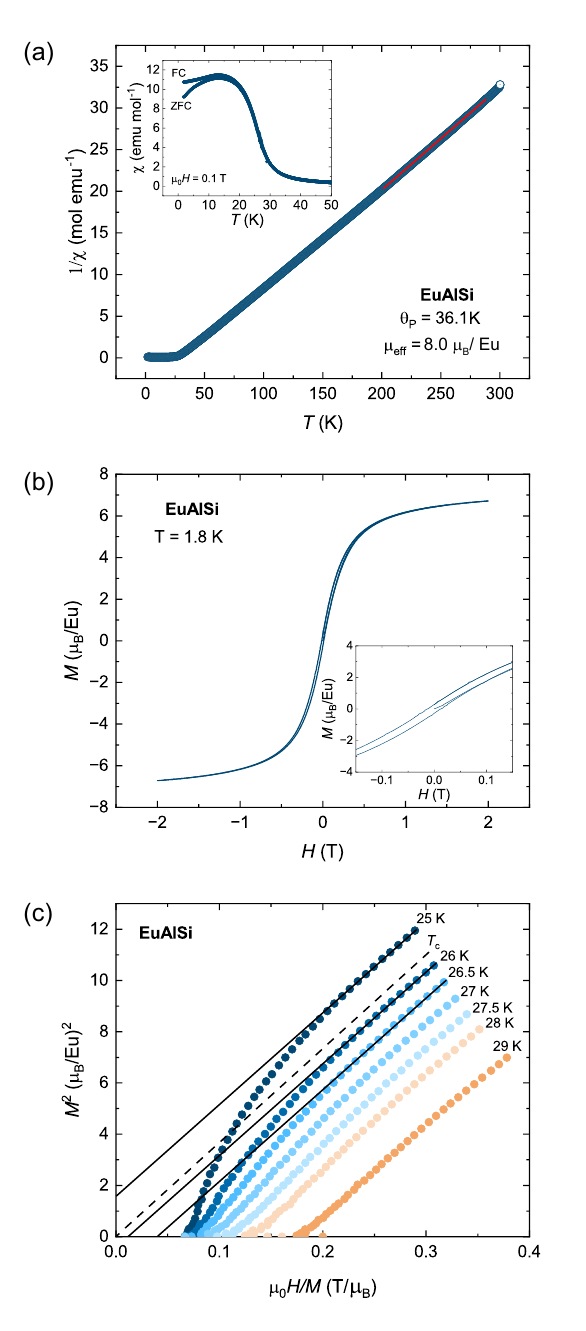}
	\caption{Magnetic measurements of EuAlSi. (a) Inverse magnetic susceptibility with a fit to the Curie-Weiss equation. Inset: ZFC and FC magnetic susceptibility in magnetic field of 0.1 T. (b) Hysteresis loop measured at 1.8 K. (c) Arrot plot, which is a plot of the square of the magnetization $M^2$ vs. the ratio of the applied magnetic field to magnetization $H/M$ at fixed temperatures between 25 K and 29 K.}
	\label{fig:fig2}
\end{figure}

\begin{table}
\caption{Details of the SXRD measurements and structural refinement for EuAlSi.}
\begin{singlespace}
    \centering
    \begin{tabular}{cc}
        \hline
        \hline
 		\multicolumn{2}{c}{Single crystal data for EuAlSi} \\
 		\hline
Composition & EuAlSi \\
CCDC/FIZ   & 2384678 \\
Formula weight [g/mol] & 207.03  \\
Temperature [K] &	160.0(1)  \\
Crystal system & hexagonal \\
Space group & \textit{P}6/\textit{mmm} \\
\textit{a} [\r{A}] & 4.2229(10)  \\
\textit{b} [\r{A}] & 4.2229(10)  \\
\textit{c} [\r{A}] & 4.5268(12)  \\
$\alpha$ [\degree] & 90 \\
$\beta$ [\degree] & 90 \\
$\gamma$ [\degree] & 120 \\
Volume [\r{A}\textsuperscript{3}] & 69.91(4) \\
\textit{Z} &	1 \\
$\rho$\textsubscript{calc} [g/cm\textsuperscript{3}] & 4.917 \\
$\mu$ [mm\textsuperscript{-1}] & 22.799 \\
\textit{F}(000) & 90.0  \\
Crystal size [mm\textsuperscript{3}] &  0.13 × 0.08 × 0.06  \\
Radiation & Mo-K$\alpha$ ($\lambda$ = 0.71073 \r{A}) \\
2$\theta$ range for data collection [\degree] & 9.004 to 60.65 \\
Index ranges & -4 $\leq$ $h$ $\leq$ 5 \\
& -6 $\leq$ $k$ $\leq$ 5 \\
& -6 $\leq$ $l$ $\leq$ 6 \\
Reflections collected &	489 \\
Independent reflections & 63 [\textit{R}\textsubscript{int} = 0.0327 \\
& \textit{R}\textsubscript{sigma} = 0.0127] \\
Data/restraints/parameters & 63/0/5  \\
Goodness-of-fit on \textit{F}\textsuperscript{2} & 1.207  \\
Final \textit{R} indexes [I $\geq$ 2 $\sigma$ (I)] & \textit{R}\textsubscript{1} =  0.0143 \\
& \textit{wR}\textsubscript{2} = 0.0356 \\
Final \textit{R} indexes [all data] & \textit{R}\textsubscript{1} = 0.0143 \\
& \textit{wR}\textsubscript{2} = 0.0356  \\
Largest diff. peak/hole [eÅ\textsuperscript{-3}] & 1.12/-1.08  \\
\hline
\hline
    \end{tabular}
    \label{tab:crystal}
    \end{singlespace}
\end{table}
   

 \begin{table}
 \caption{Atomic coordinates, isotropic displacement parameters, and occupancy of the atoms in EuAlSi obtained by SXRD.}
     \centering
     \begin{tabular}{ccccccc}
     \hline
     \hline
 		\multicolumn{7}{c}{Atomic positions} \\
 		\hline
 		Atom & Symbol & $x$ & $y$ & $z$ & U\textsubscript{iso} & Occ. \\ \hline
            Eu & \textit{1a} & 0 & 1 & 1 & 0.0058(2) & 1 \\
 		Al & \textit{2d} & 1/3 & 2/3 & 1/2 & 0.0097(5) & 0.5\\
 		Si & \textit{2d} & 1/3 & 2/3 & 1/2 & 0.0097(5) & 0.5\\	
 	\hline
 	\hline
    \end{tabular}
    \label{tab:Wyckoff}
 \end{table}
 

\subsection{Magnetic properties of EuAlSi}

\begin{figure}
\centering
	\includegraphics[width=1\linewidth]{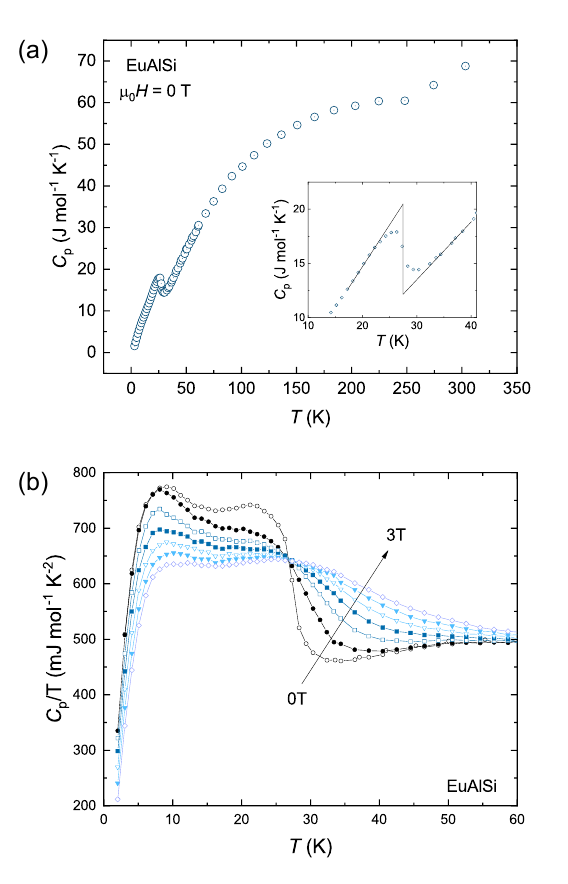}
	\caption{Heat capacity of EuAlSi (a) measured in zero magnetic field, (b) measured in external magnetic field ranging from 0 T to 3 T in 0.5 T steps.}
	\label{fig:fig3}
\end{figure}

The inset in Figure \ref{fig:fig2}(a) shows the zero field cooled (ZFC) and field cooled (FC) magnetic susceptibility, here defined as $\chi = M/H$, of EuAlSi measured in an external magnetic field of $\mu_0 H =$ 0.1 T in the temperature range between $T =$ 1.8 and 50 K. The sudden increase in the magnetic susceptibility around 30 K and saturation of the signal in the FC plot suggest a transition into a long-range ferromagnetic ordered state. To investigate the nature of the magnetic ordering in EuAlSi, the inverse of magnetic susceptibility 1/$\chi$ (shown in Figure \ref{fig:fig2}(a)) in the high-temperature region ($T$ = 200--290 K) was analyzed using the Curie-Weiss law:  

\begin{equation}
    \chi(T) = \frac{C}{T-\theta_P},
     \label{eq:6C-W}
\end{equation}

where $C$ is the Curie constant and $\theta_P$ is the paramagnetic Curie temperature. From the fit, shown in Figure \ref{fig:fig2}(a), we obtained $C$ = 8 emu K and $\theta_P$ = 36.1 K. The positive value of the paramagnetic Curie temperature agrees well with the presence of nearest-neighbor ferromagnetic interactions in the paramagnetic state. The effective magnetic moment of Eu was calculated as $C = {\mu^2_{eff}}/8$ and the obtained value of $\mu_{eff}$ = 8.0 $\mu_B$/Eu is in excellent agreement with the theoretical value of 7.9 $\mu_B$ for Eu$^{2+}$ free ions. Additionally, to the data shown in the inset of Figure \ref{fig:fig2}(a), ZFC and FC magnetic susceptibility measurements in magnetic field of $\mu_0 H =$ 0.01 T and 1 T are shown in Figure S2 in the Supplementary Material. All measurements show the same increase in the magnetization around 30 K and the saturation of the magnetic moment below the transition temperature. The value of the transition temperature, $T_{\rm Curie}$, estimated as a minimum of $dM/dT$ for $\mu_0 H$ = 0.1 T is equal to $T_{\rm Curie}$ = 26.8 K. The visible difference in ZFC and FC measurements at low fields is, as expected, caused by the domain wall pinning, as a consequence of the magneto-crystalline anisotropy \cite{Salamakha2013, Tian2019, Gornicka2019}. The difference in ZFC and FC measurements disappear for magnetic fields of $\mu_0 H$ $>$ 1 T, due to fully saturated magnetization and all magnetic domains are alligned in these fields.

Field-dependent magnetization $M(H)$ measurements at $T$ = 1.8 K are shown in Figure \ref{fig:fig2}(b). The small hysteresis loop indicates that EuAlSi is a soft ferromagnetic material with the coercivity field of only $\approx$130 G and a remnant magnetization of 0.25 $\mu_B$/Eu, which is 3.7 \% of the magnetization obtained at 2 T. Our measurements of $M(H)$ at $T$ = 10 and 40 K are shown in Figure S1 in the SM. The hysteresis loop measured at $T$ = 10 K displays a coercivity field and remanent magnetization close to zero, while at 40 K, the $M(H)$ curve is a straight line corresponding to the paramagnetic state of EuAlSi.

An Arrot plot is presented in Figure \ref{fig:fig2}(c). The magnetic isotherms were measured in the vicinity of the transition temperature and are plotted as $M^2$ versus $H/M$. This representation allows to precisely determine $T_{\rm Curie}$ \cite{Arrot1957,Lefvre2022}. Thereby, the isotherm at the Curie temperature will pass through the origin, as indicated by the straight lines in Figure \ref{fig:fig2}(c). In Figure \ref{fig:fig2}(c) the isotherm measured at 25 K shows clear downturns, which might be caused by the presence of magnetocrystalline anisotropy \cite{Pramanik2009, Salamakha2013}. The straight line that goes through the origin lies in between 25 and 26 K, indicating that the Curie temperature is close to 26 K (see Figure \ref{fig:fig2}(c)), while the isotherms gathered at higher temperatures have a form of straight lines, indicating the validity of the mean-field approximation \cite{Pramanik2009}. The $T_{\rm Curie}$ assigned from the Arrot plot is in agreement with the $T_{\rm Curie}$ assigned from the first derivative $d\chi/dT$, which was found to be $T_{\rm Curie}$ = 26.8 K.

\begin{figure}
\centering
\includegraphics[width=1\linewidth]{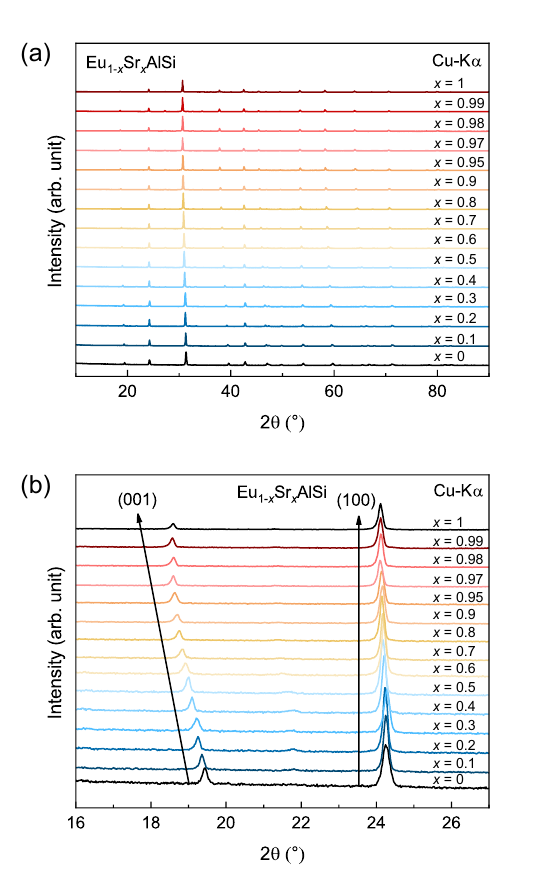}
\caption{PXRD patterns of the \ce{Eu_{1-\textit{x}}Sr_{\textit{x}}AlSi} solid solution (a) in a 2$\theta$ range between 10 and 90$\degree$, (b) in the 2$\theta$ range between 16 and 27$\degree$ for a better visibility of the evolution of the (100) and the (001) reflections.}

\label{fig:fig4}
\end{figure}

To further characterize the magnetic order in EuAlSi the heat capacity of this material was measured. The heat capacity in zero magnetic field is shown in Figure \ref{fig:fig3}(a). The measured data at high temperature, in agreement with Dulong-Petit law, reaches the limit of $3nR$ =  74.49 J mol$^{-1}$ K$^{-1}$, where $n$ is the number of atoms and $R$ is the ideal gas constant. In the low-temperature region we observe, as expected, a jump in the heat capacity at the phase transition to the long-range ferromagnetic state. In the inset of Figure \ref{fig:fig3}(a) we show the entropy conserving geometrical construction, which allows us to assign the thermodynamic critical temperature of the transition to the ferromagnetic state, which was found to be $T_{\rm Curie}$ = 27.5 K. The $T_{\rm Curie}$ assigned from heat capacity is in agreement with $T_{Curie}$ assigned from $d\chi/dT$ and Arrott plots. The large value of $\Delta C_p$ confirms the bulk nature of the long-range magnetic order in EuAlSi.

Figure \ref{fig:fig3}(b) presents the temperature dependence of $C_p/T$ below 80 K with field from 0 T to 3 T. Two maxima of $C/T$ are observed. The maximum at $\approx$ 27 K is the transition to the ferromagnetic state seen in $C_p$ in Figure \ref{fig:fig3}(a), while the maximum at $\approx$ 10 K is a Schottky anomaly, observed in Eu$^{2+}$ compounds \cite{fu2020, Adhikari2019, Kaczorowski2012, Kumar2011}, and which reflects thermal variation in the population of split levels of the Eu$^{2+}$ ground state \cite{Tari2003}. Both maxima gradually disappear with increasing magnetic field.

\subsection{Crystal structure of the \ce{Eu_{1-\textit{x}}Sr_{\textit{x}}AlSi} solid solution}\label{sec:solid-solution}

\begin{figure}
\centering
\includegraphics[width=1\linewidth]{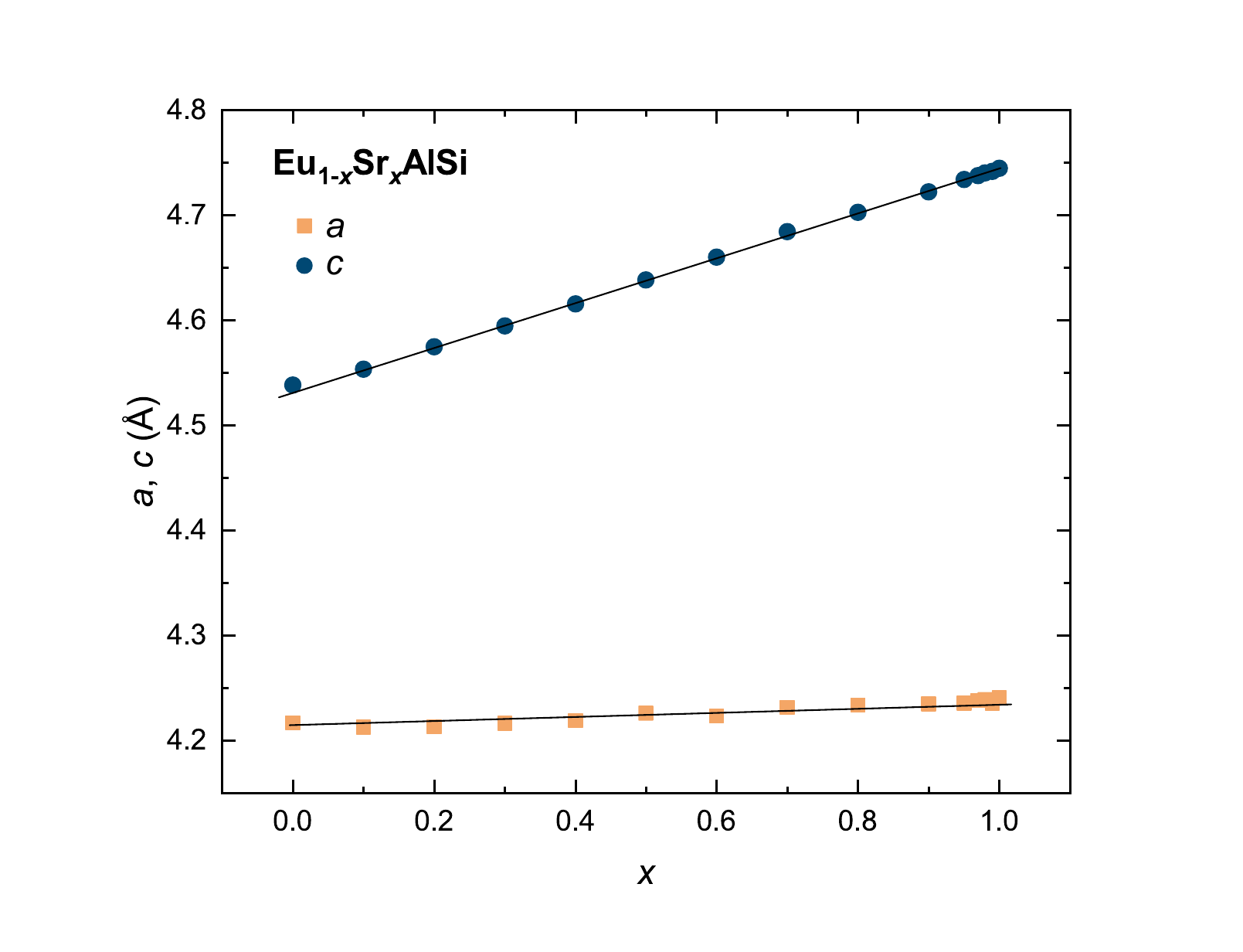}
\caption{Unit cell parameters of all members of the \ce{Eu_{1-\textit{x}}Sr_{\textit{x}}AlSi} solid solution. Black line is a guide for the eye.}
\label{fig:fig5}
\end{figure}

We synthesized the \ce{Eu_{1-\textit{x}}Sr_{\textit{x}}AlSi} solid solution, transitioning from the ferromagnetic Eu-rich end member to the superconducting SrAlSi end member by systematically substituting Eu$^{2+}$ with the isovalent Sr$^{2+}$. SrAlSi crystallizes in the same structure type as EuAlSi in the \textit{P}6/\textit{mmm} space group \cite{Evans2009, Walicka2021}. All members of the \ce{Eu_{1-\textit{x}}Sr_{\textit{x}}AlSi} solid solution were prepared by arc melting and their crystal structure and phase purity were analyzed by PXRD. All PXRD patterns are presented in Figure \ref{fig:fig4}(a) showing that all members of the solid solution crystallize in the \textit{P}6/\textit{mmm} space group as their parent compounds. In Figure \ref{fig:fig4}(b) the evolution of the two first Bragg reflections is presented: (001) and (100), respectively. As one can see, the (001) peak is shifting towards lower angles, indicating a systematic increase in the $c$ parameter, while the position of the (100) peak remains almost constant. The unit cell parameters extracted from the PXRD by LeBail refinement, presented in Figure \ref{fig:fig5}, show almost no change in the cell parameter $a$, while the cell parameter $c$ increases significantly across the solid solution for increasing Sr contents. Both parameters change linearly in agreement with Vegard's law, suggesting that the substitution of Eu and Sr in \ce{Eu_{1-\textit{x}}Sr_{\textit{x}}AlSi} solid solution is homogeneous and without phase separation or structural disruption. Preservation of the hexagonal crystal structure across the solid solution suggest that changes in physical properties are more likely intrinsic, and not arising from defects or inhomogeneities. This substitutional stability makes the system reliable for studying composition-dependence of the interplay between the structure and physical properties.

\begin{figure*}
\centering
	\includegraphics[width=1\linewidth]{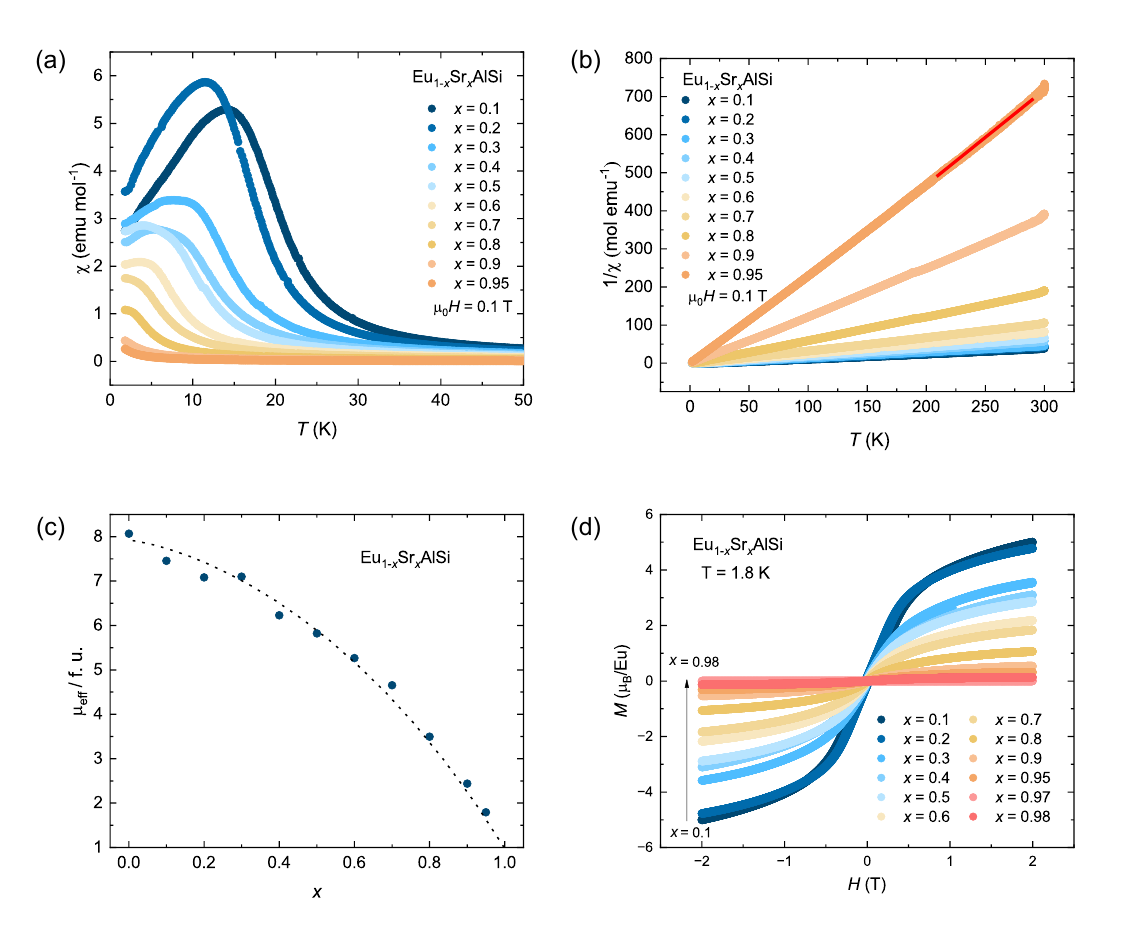}
	\caption{(a) Magnetic susceptibility of all magnetic samples of the \ce{Eu_{1-\textit{x}}Sr_{\textit{x}}AlSi} solid solution ($0.1 \leq x \leq 0.95$). (b) Inverse of the magnetic susceptibility of samples with $0.1 \leq x \leq 0.95$ together with Curie-Weiss fit. (c) Effective magnetic moment per formula unit for all magnetic members of the solid solution ($0.1 \leq x \leq 0.95$). (d) Hysteresis loop for all the magnetic members of the \ce{Eu_{1-\textit{x}}Sr_{\textit{x}}AlSi} solid solution.}
	\label{fig:fig6}
\end{figure*}
\begin{figure}
\centering
	\includegraphics[width=1\linewidth]{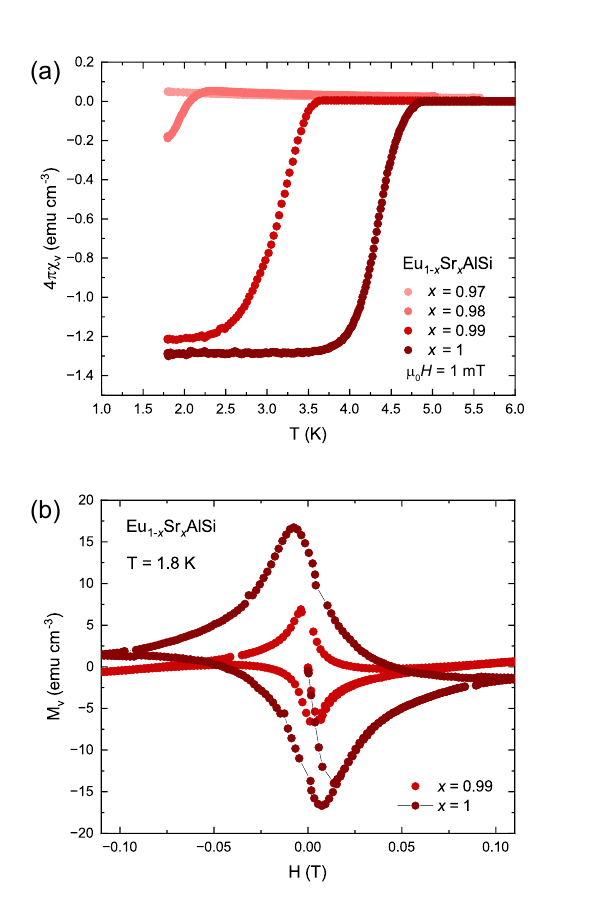}
	\caption{(a) Volumetric susceptibility for samples with superconducting contribution, corrected for the demagnetization field. (b) $M(H)$ for superconducting members of the \ce{Eu_{1-\textit{x}}Sr_{\textit{x}}AlSi} solid solution. }
	\label{fig:fig7}
\end{figure}

\begin{figure}
\centering
	\includegraphics[width=1\linewidth]{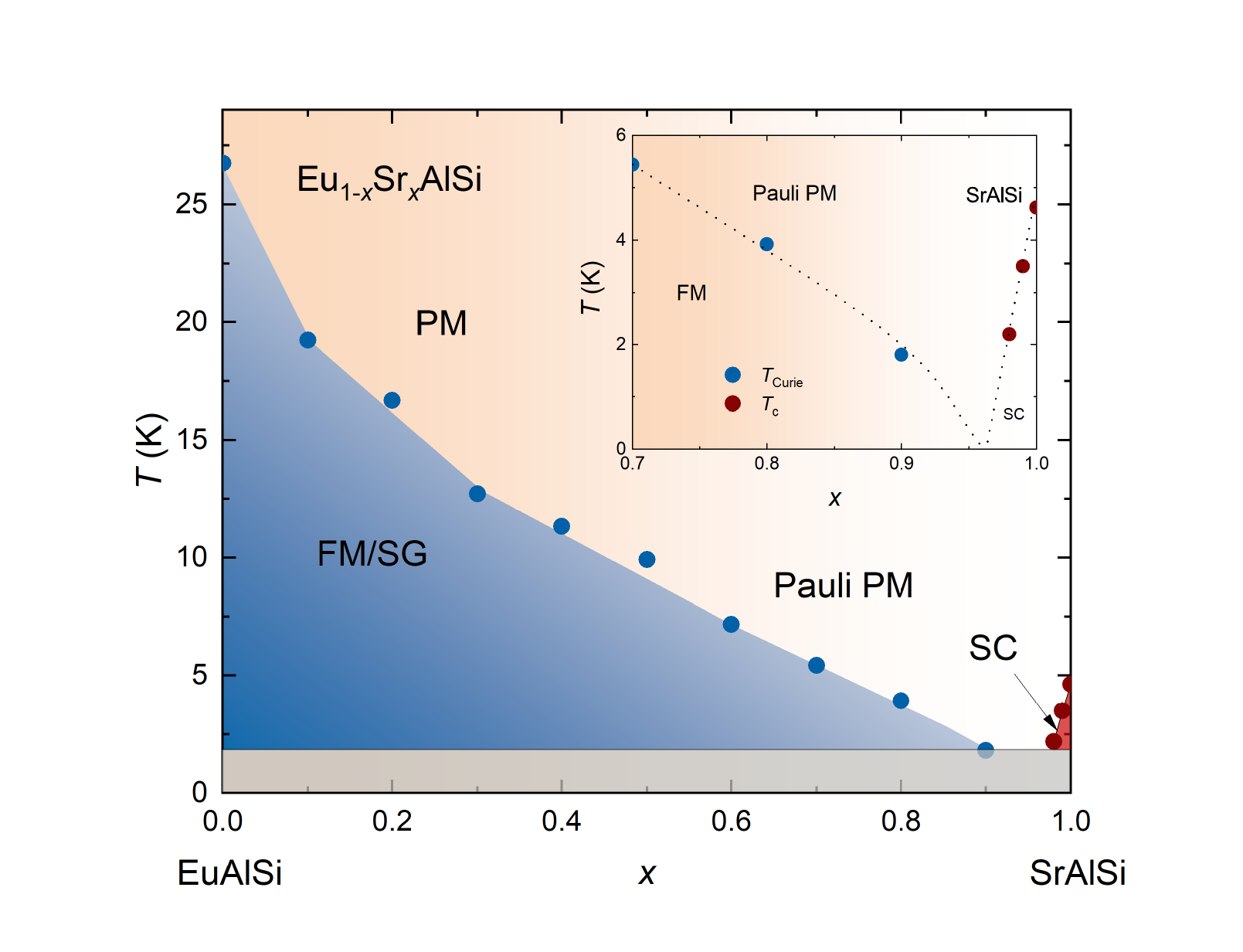}
	\caption{Phase diagram of the \ce{Eu_{1-\textit{x}}Sr_{\textit{x}}AlSi} solid solution. }
	\label{fig:fig8}
\end{figure}

\subsection{Magnetic properties of the \ce{Eu_{1-\textit{x}}Sr_{\textit{x}}AlSi} solid solution}

The temperature-dependent magnetization measured in a magnetic field of 0.1 T for samples with $0.1 \leq x \leq 0.95$ is shown in Figure \ref{fig:fig6}(a). For those samples, similar as for EuAlSi, anincrease in the magnetization is observed at low temperatures. The evolution of the signal with an increasing Sr content is clearly observed, and with increasing $x$ the onset of increasing magnetization is shifting towards lower temperatures and the saturation of the signal is reached at lower values of $\chi$ (see Figure \ref{fig:fig6}(a)). ZFC and FC measurements in applied magnetic fields of 0.01, 0.1 and 1 T are shown in Figures S2, S3 and S4 in the SM. Interestingly, for most of the measured samples, both ZFC and FC curves show a downturn after reaching the maximum value in small external magnetic fields, which might be caused by disorder, appearing upon dilution of the Eu atoms on the triangular lattice \cite{Blundell, Tian2019}. For all samples, a difference between ZFC and FC measurements is observed, caused by a domain wall pinning effect. This difference disappears with increasing magnetic field (see Figures S2, S3 and S4 in the SM). The values of $T_{\rm Curie}$ for all magnetic samples were assigned as the minimum in the $d\chi/dT$ of the $M(T)$ data as measured in a magnetic field of 0.1 T, and are shown in the Table S2 in the SM and Figure \ref{fig:fig8}. The value of $T_{\rm Curie}$ decreases with increasing $x$ as expected. 

The inverse of the magnetic susceptibility 1/$\chi$ together with the fit to the Curie-Weiss equation \ref{eq:6C-W} is shown in Figure \ref{fig:fig6}(b). The obtained Curie constant $C$ and paramagnetic Curie temperature $\theta_P$ for all members of the solid solution are shown in Table S2 in the SM, while the effective magnetic moments are presented in Table S2 in the SM and in Figure \ref{fig:fig6}(c). As expected, the substitution of Eu for Sr in \ce{Eu_{1-\textit{x}}Sr_{\textit{x}}AlSi} (increase in $x$) causes a decrease of the effective magnetic moment $\mu_{\rm eff}$. The $\mu_{\rm eff}$ decreases from 8.0 $\mu_B$/f.u. for EuAlSi to 1.8 $\mu_B$/f.u. for sample with nominal composition \ce{Eu_{0.05}Sr_{0.95}AlSi}.

The $M(H)$ measurements for all magnetic members of the solid solution are presented in Figure \ref{fig:fig6}(d). The hysteresis loops show systematically lower saturation value at the magnetic field of 2 T, confirming smaller values of the effective magnetic moments with increasing $x$. All the curves show a characteristic S-shape, which might not be obvious from the scale of the Figure, hence the $M(H)$ curves for $x$ = 0.97 and 0.98 are additionally shown in Figure S4 in SM.

Temperature dependence of the volumetric susceptibility for the four samples near \ce{SrAlSi} is presented in Figure \ref{fig:fig7}(a). The sample with a nominal $x$ = 0.97 is a paramgnet down to 1.8 K, while the one with $x$ = 0.98 shows a downturn of the susceptibility at $T_{\rm c}$ = 2 K indicating presence of the superconducting state. A much stronger diamagnetic signal and a saturation of $\chi_v$ below the value of -1 is observed for $x$ = 0.99 and 1.0. The data for these two samples were corrected for the demagnetization factor, as commonly done for the superconducting materials \cite{klimczuk2023}. The superconducting temperatures are $T_{\rm c}$ = 3.5 K and 4.6 K, for $x$ = 0.99 and 1.0, respectively. The $M(H)$ measurements in the superconducting state at $T$ = 1.8 K are shown in Figure \ref{fig:fig7}(b).  

A summary of the magnetic and superconducting properties of the \ce{Eu_{1-\textit{x}}Sr_{\textit{x}}AlSi} solid solution is shown in Figure \ref{fig:fig8}. The gray bar at the bottom of the figure indicates the temperature range, which we are not able to access. In the presented phase diagram, the sample with $x$ = 0 is ferromagnetic. Samples with $0.1 \leq x \leq 0.9$ are magnetic, with systematically decreasing $T_{\rm Curie}$ from 26.8 K for EuAlSi to 1.8 K for \ce{Eu_{0.1}Sr_{0.9}AlSi} in a linear way. The ground state of these samples hints towards a spin glass behavior. The inset magnifies the phase diagram near pure \ce{SrAlSi} and the dashed lines are guides to eyes. Both lines could be extrapolated to 0 K at $x$ around 0.96, which might suggest the existence of a potential quantum critical point (QCP) above which the ferromagnetic Curie temperature vanishes, and superconductivity emerges.
The nearly linear suppression of FM $T_{\rm Curie}$ with $x$ may suggest that quantum fluctuations become dominant at very low temperatures. Probing transport, specific heat, and magnetic fluctuations (i.e. inelastic neutron scattering) near $x$ = 0.96 may thus reveal the non-Fermi liquid behavior that is a hallmark of quantum criticality.

\section{Summary and conclusion}
\label{sec:conclusion6}

We have successfully synthesized EuAlSi and in detail analyzed its structural and magnetic properties. By means of single-crystal X-ray diffraction we found that EuAlSi crystallizes into a \ce{AlB2}-type structure in the \textit{P}6/\textit{mmm} space group with unit cell parameters $a$ = 4.2229(10) \r{A} and $c$ = 4.5268(12) \r{A}. Using magnetization measurements, we have shown that this compound is a soft ferromagnetic material with $T_{Curie}$ = 25.8 K. We have obtained a paramagnetic Curie temperature of $\theta_P$ = 36.1 K and an effective magnetic moment $\mu_{eff}$ = 8.0 $\mu_B$/Eu, which is in very good agreement with the theoretical value of 7.9 $\mu_B$ for Eu$^{2+}$ free ion. The presence of the heat capacity jump at 27.5 K, confirms the bulk nature of the ferromagnetic transition.

Moreover, we have successfully synthesized the \ce{Eu_{1-\textit{x}}Sr_{\textit{x}}AlSi} solid solution to study the interplay between ferromagnetism and superconductivity in this system. We found that all samples of the solid solution crystallize in the space group \textit{P}6/\textit{mmm} and that the unit cell parameters across the \ce{Eu_{1-\textit{x}}Sr_{\textit{x}}AlSi} solid solution change in a linear fashion in agreement with Vegard's law. By analyzing the magnetic properties, we were able to obtain a comprehensive ground state phase diagram of the \ce{Eu_{1-\textit{x}}Sr_{\textit{x}}AlSi} solid solution down to 1.8 K. We showed that $T_{\rm Curie}$ systematically decreases from 26.8 K for EuAlSi to 1.8 K for \ce{Eu_{0.1}Sr_{0.9}AlSi} in a linear fashion and that superconductivity is fully suppressed already with an Eu content of 0.05 Eu ($x$ = 0.95). Above the critical temperatures there is a transition from paramagnetic state to temperature-independent paramagnetic state with decreasing $x$ content. The superconducting state in this system is observed to occur within a very narrow range. Even the smallest concentration (3 \%) of Eu has been demonstrated to effectively break up the Cooper pairs and destroy the superconducting state, which aligns with the expected behavior of a singlet-pairing state in SrAlSi that is rapidly suppressed in the presence of magnetic impurities. It is noteworthy that even lower concentration value (1 \%) of magnetic gadolinium is required to suppress the superconducting state in pure lanthanum \cite{Matthias1958}. A possible quantum critical point is suggested to exist in the vicinity of $x$~0.96, at which the suppression of ferromagnetic order is concomitant with the emergence of superconductivity, and a predominance of quantum fluctuations may possibly stabilize more exotic phenomena such a non-Fermi liquid resistivity and unconventional superconducting pairing. Further investigation at ultra-low temperatures is essential to elucidate the details of the putative quantum critical behavior, and its role on the interplay between magnetism and superconductivity.

\section{Acknowledgment}
\label{sec:Acknowledgement}
This work was supported  by the Swiss National Science Foundation under Grants No. PCEFP2\_194183.

\bibliography{bib}

\begin{thebibliography}{35}%
\makeatletter
\providecommand \@ifxundefined [1]{%
 \@ifx{#1\undefined}
}%
\providecommand \@ifnum [1]{%
 \ifnum #1\expandafter \@firstoftwo
 \else \expandafter \@secondoftwo
 \fi
}%
\providecommand \@ifx [1]{%
 \ifx #1\expandafter \@firstoftwo
 \else \expandafter \@secondoftwo
 \fi
}%
\providecommand \natexlab [1]{#1}%
\providecommand \enquote  [1]{``#1''}%
\providecommand \bibnamefont  [1]{#1}%
\providecommand \bibfnamefont [1]{#1}%
\providecommand \citenamefont [1]{#1}%
\providecommand \href@noop [0]{\@secondoftwo}%
\providecommand \href [0]{\begingroup \@sanitize@url \@href}%
\providecommand \@href[1]{\@@startlink{#1}\@@href}%
\providecommand \@@href[1]{\endgroup#1\@@endlink}%
\providecommand \@sanitize@url [0]{\catcode `\\12\catcode `\$12\catcode `\&12\catcode `\#12\catcode `\^12\catcode `\_12\catcode `\%12\relax}%
\providecommand \@@startlink[1]{}%
\providecommand \@@endlink[0]{}%
\providecommand \url  [0]{\begingroup\@sanitize@url \@url }%
\providecommand \@url [1]{\endgroup\@href {#1}{\urlprefix }}%
\providecommand \urlprefix  [0]{URL }%
\providecommand \Eprint [0]{\href }%
\providecommand \doibase [0]{https://doi.org/}%
\providecommand \selectlanguage [0]{\@gobble}%
\providecommand \bibinfo  [0]{\@secondoftwo}%
\providecommand \bibfield  [0]{\@secondoftwo}%
\providecommand \translation [1]{[#1]}%
\providecommand \BibitemOpen [0]{}%
\providecommand \bibitemStop [0]{}%
\providecommand \bibitemNoStop [0]{.\EOS\space}%
\providecommand \EOS [0]{\spacefactor3000\relax}%
\providecommand \BibitemShut  [1]{\csname bibitem#1\endcsname}%
\let\auto@bib@innerbib\@empty
\bibitem [{\citenamefont {Cao}\ \emph {et~al.}(2018)\citenamefont {Cao}, \citenamefont {Fatemi}, \citenamefont {Fang}, \citenamefont {Watanabe}, \citenamefont {Taniguchi}, \citenamefont {Kaxiras},\ and\ \citenamefont {Jarillo-Herrero}}]{Cao2018}%
  \BibitemOpen
  \bibfield  {author} {\bibinfo {author} {\bibfnamefont {Y.}~\bibnamefont {Cao}}, \bibinfo {author} {\bibfnamefont {V.}~\bibnamefont {Fatemi}}, \bibinfo {author} {\bibfnamefont {S.}~\bibnamefont {Fang}}, \bibinfo {author} {\bibfnamefont {K.}~\bibnamefont {Watanabe}}, \bibinfo {author} {\bibfnamefont {T.}~\bibnamefont {Taniguchi}}, \bibinfo {author} {\bibfnamefont {E.}~\bibnamefont {Kaxiras}},\ and\ \bibinfo {author} {\bibfnamefont {P.}~\bibnamefont {Jarillo-Herrero}},\ }\bibfield  {title} {\bibinfo {title} {Unconventional superconductivity in magic-angle graphene superlattices},\ }\href@noop {} {\bibfield  {journal} {\bibinfo  {journal} {Nature}\ }\textbf {\bibinfo {volume} {556}},\ \bibinfo {pages} {43} (\bibinfo {year} {2018})}\BibitemShut {NoStop}%
\bibitem [{\citenamefont {Nagamatsu}\ \emph {et~al.}(2001)\citenamefont {Nagamatsu}, \citenamefont {Nakagawa}, \citenamefont {Muranaka}, \citenamefont {Zenitani},\ and\ \citenamefont {Akimitsu}}]{Nagamatsu2001}%
  \BibitemOpen
  \bibfield  {author} {\bibinfo {author} {\bibfnamefont {J.}~\bibnamefont {Nagamatsu}}, \bibinfo {author} {\bibfnamefont {N.}~\bibnamefont {Nakagawa}}, \bibinfo {author} {\bibfnamefont {T.}~\bibnamefont {Muranaka}}, \bibinfo {author} {\bibfnamefont {Y.}~\bibnamefont {Zenitani}},\ and\ \bibinfo {author} {\bibfnamefont {J.}~\bibnamefont {Akimitsu}},\ }\bibfield  {title} {\bibinfo {title} {{Superconductivity at 39 K in magnesium diboride}},\ }\href {https://doi.org/10.1038/35065039} {\bibfield  {journal} {\bibinfo  {journal} {Nature}\ }\textbf {\bibinfo {volume} {410}},\ \bibinfo {pages} {63} (\bibinfo {year} {2001})}\BibitemShut {NoStop}%
\bibitem [{\citenamefont {Evans}\ \emph {et~al.}(2009)\citenamefont {Evans}, \citenamefont {Wu}, \citenamefont {Kranak}, \citenamefont {Newman}, \citenamefont {Reller}, \citenamefont {Garcia-Garcia},\ and\ \citenamefont {H{\"{a}}ussermann}}]{Evans2009}%
  \BibitemOpen
  \bibfield  {author} {\bibinfo {author} {\bibfnamefont {M.~J.}\ \bibnamefont {Evans}}, \bibinfo {author} {\bibfnamefont {Y.}~\bibnamefont {Wu}}, \bibinfo {author} {\bibfnamefont {V.~F.}\ \bibnamefont {Kranak}}, \bibinfo {author} {\bibfnamefont {N.}~\bibnamefont {Newman}}, \bibinfo {author} {\bibfnamefont {A.}~\bibnamefont {Reller}}, \bibinfo {author} {\bibfnamefont {F.~J.}\ \bibnamefont {Garcia-Garcia}},\ and\ \bibinfo {author} {\bibfnamefont {U.}~\bibnamefont {H{\"{a}}ussermann}},\ }\bibfield  {title} {\bibinfo {title} {{Structural properties and superconductivity in the ternary intermetallic compounds MAB (M=Ca, Sr, Ba; A=Al, Ga, In; B=Si, Ge, Sn)}},\ }\href {https://doi.org/10.1103/PhysRevB.80.064514} {\bibfield  {journal} {\bibinfo  {journal} {Phys. Rev. B}\ }\textbf {\bibinfo {volume} {80}},\ \bibinfo {pages} {064514} (\bibinfo {year} {2009})}\BibitemShut {NoStop}%
\bibitem [{\citenamefont {Walicka}\ \emph {et~al.}(2021)\citenamefont {Walicka}, \citenamefont {Guguchia}, \citenamefont {Lago}, \citenamefont {Blacque}, \citenamefont {Ma}, \citenamefont {Liu}, \citenamefont {Khasanov},\ and\ \citenamefont {von Rohr}}]{Walicka2021}%
  \BibitemOpen
  \bibfield  {author} {\bibinfo {author} {\bibfnamefont {D.~I.}\ \bibnamefont {Walicka}}, \bibinfo {author} {\bibfnamefont {Z.}~\bibnamefont {Guguchia}}, \bibinfo {author} {\bibfnamefont {J.}~\bibnamefont {Lago}}, \bibinfo {author} {\bibfnamefont {O.}~\bibnamefont {Blacque}}, \bibinfo {author} {\bibfnamefont {K.}~\bibnamefont {Ma}}, \bibinfo {author} {\bibfnamefont {H.}~\bibnamefont {Liu}}, \bibinfo {author} {\bibfnamefont {R.}~\bibnamefont {Khasanov}},\ and\ \bibinfo {author} {\bibfnamefont {F.~O.}\ \bibnamefont {von Rohr}},\ }\bibfield  {title} {\bibinfo {title} {Two-gap to single-gap superconducting transition on a honeycomb lattice in \ce{Ca_{1-x}Sr_{x}AlSi}},\ }\href {https://doi.org/10.1103/PhysRevResearch.3.033192} {\bibfield  {journal} {\bibinfo  {journal} {Phys. Rev. Research}\ }\textbf {\bibinfo {volume} {3}},\ \bibinfo {pages} {033192} (\bibinfo {year} {2021})}\BibitemShut {NoStop}%
\bibitem [{\citenamefont {Tran}\ \emph {et~al.}(2021)\citenamefont {Tran}, \citenamefont {Sahakyan},\ and\ \citenamefont {Bukowski}}]{Tran2021}%
  \BibitemOpen
  \bibfield  {author} {\bibinfo {author} {\bibfnamefont {V.~H.}\ \bibnamefont {Tran}}, \bibinfo {author} {\bibfnamefont {M.}~\bibnamefont {Sahakyan}},\ and\ \bibinfo {author} {\bibfnamefont {Z.}~\bibnamefont {Bukowski}},\ }\bibfield  {title} {\bibinfo {title} {Discovery of superconductivity in \ce{AlB2}-type hexagonal \ce{YGa2}},\ }\href@noop {} {\bibfield  {journal} {\bibinfo  {journal} {Journal of Physics: Condensed Matter}\ }\textbf {\bibinfo {volume} {33}},\ \bibinfo {pages} {315401} (\bibinfo {year} {2021})}\BibitemShut {NoStop}%
\bibitem [{\citenamefont {Walicka}\ \emph {et~al.}(2023)\citenamefont {Walicka}, \citenamefont {Lef\`evre}, \citenamefont {Blacque}, \citenamefont {L\'opez-Paz}, \citenamefont {Rischau}, \citenamefont {Cervellino}, \citenamefont {Triana},\ and\ \citenamefont {von Rohr}}]{walicka2023bagage}%
  \BibitemOpen
  \bibfield  {author} {\bibinfo {author} {\bibfnamefont {D.~I.}\ \bibnamefont {Walicka}}, \bibinfo {author} {\bibfnamefont {R.}~\bibnamefont {Lef\`evre}}, \bibinfo {author} {\bibfnamefont {O.}~\bibnamefont {Blacque}}, \bibinfo {author} {\bibfnamefont {S.~A.}\ \bibnamefont {L\'opez-Paz}}, \bibinfo {author} {\bibfnamefont {C.~W.}\ \bibnamefont {Rischau}}, \bibinfo {author} {\bibfnamefont {A.}~\bibnamefont {Cervellino}}, \bibinfo {author} {\bibfnamefont {C.~A.}\ \bibnamefont {Triana}},\ and\ \bibinfo {author} {\bibfnamefont {F.~O.}\ \bibnamefont {von Rohr}},\ }\bibfield  {title} {\bibinfo {title} {Structural phase transition and superconductivity in {2H-BaGaGe} with buckled honeycomb layers},\ }\href@noop {} {\bibfield  {journal} {\bibinfo  {journal} {Phys. Rev. Mater.}\ }\textbf {\bibinfo {volume} {7}},\ \bibinfo {pages} {074805} (\bibinfo {year} {2023})}\BibitemShut {NoStop}%
\bibitem [{\citenamefont {Walicka}\ \emph {et~al.}(2024)\citenamefont {Walicka}, \citenamefont {Blacque}, \citenamefont {Klimczuk},\ and\ \citenamefont {von Rohr}}]{Walicka2024}%
  \BibitemOpen
  \bibfield  {author} {\bibinfo {author} {\bibfnamefont {D.~I.}\ \bibnamefont {Walicka}}, \bibinfo {author} {\bibfnamefont {O.}~\bibnamefont {Blacque}}, \bibinfo {author} {\bibfnamefont {T.}~\bibnamefont {Klimczuk}},\ and\ \bibinfo {author} {\bibfnamefont {F.~O.}\ \bibnamefont {von Rohr}},\ }\bibfield  {title} {\bibinfo {title} {From weak- to strong-coupling superconductivity in the \ce{AlB2}-type solid solution \ce{SrGa_{1--x}Al_{x}Ge} with honeycomb layers},\ }\href@noop {} {\bibfield  {journal} {\bibinfo  {journal} {Journal of Physics: Condensed Matter}\ }\textbf {\bibinfo {volume} {37}},\ \bibinfo {pages} {045704} (\bibinfo {year} {2024})}\BibitemShut {NoStop}%
\bibitem [{\citenamefont {Kurumaji}\ \emph {et~al.}(2019)\citenamefont {Kurumaji}, \citenamefont {Nakajima}, \citenamefont {Hirschberger}, \citenamefont {Kikkawa}, \citenamefont {Yamasaki}, \citenamefont {Sagayama}, \citenamefont {Nakao}, \citenamefont {Taguchi}, \citenamefont {hisa Arima},\ and\ \citenamefont {Tokura}}]{Kurumaji2019}%
  \BibitemOpen
  \bibfield  {author} {\bibinfo {author} {\bibfnamefont {T.}~\bibnamefont {Kurumaji}}, \bibinfo {author} {\bibfnamefont {T.}~\bibnamefont {Nakajima}}, \bibinfo {author} {\bibfnamefont {M.}~\bibnamefont {Hirschberger}}, \bibinfo {author} {\bibfnamefont {A.}~\bibnamefont {Kikkawa}}, \bibinfo {author} {\bibfnamefont {Y.}~\bibnamefont {Yamasaki}}, \bibinfo {author} {\bibfnamefont {H.}~\bibnamefont {Sagayama}}, \bibinfo {author} {\bibfnamefont {H.}~\bibnamefont {Nakao}}, \bibinfo {author} {\bibfnamefont {Y.}~\bibnamefont {Taguchi}}, \bibinfo {author} {\bibfnamefont {T.}~\bibnamefont {hisa Arima}},\ and\ \bibinfo {author} {\bibfnamefont {Y.}~\bibnamefont {Tokura}},\ }\bibfield  {title} {\bibinfo {title} {Skyrmion lattice with a giant topological hall effect in a frustrated triangular-lattice magnet},\ }\href {https://doi.org/10.1126/science.aau0968} {\bibfield  {journal} {\bibinfo  {journal} {Science}\ }\textbf {\bibinfo {volume} {365}},\ \bibinfo {pages} {914} (\bibinfo {year} {2019})}\BibitemShut {NoStop}%
\bibitem [{\citenamefont {Pan}\ \emph {et~al.}(2013)\citenamefont {Pan}, \citenamefont {Cao}, \citenamefont {Bai}, \citenamefont {Song}, \citenamefont {Zheng},\ and\ \citenamefont {Duan}}]{Pan2013}%
  \BibitemOpen
  \bibfield  {author} {\bibinfo {author} {\bibfnamefont {Z.-Y.}\ \bibnamefont {Pan}}, \bibinfo {author} {\bibfnamefont {C.-D.}\ \bibnamefont {Cao}}, \bibinfo {author} {\bibfnamefont {X.-J.}\ \bibnamefont {Bai}}, \bibinfo {author} {\bibfnamefont {R.-B.}\ \bibnamefont {Song}}, \bibinfo {author} {\bibfnamefont {J.-B.}\ \bibnamefont {Zheng}},\ and\ \bibinfo {author} {\bibfnamefont {L.-B.}\ \bibnamefont {Duan}},\ }\bibfield  {title} {\bibinfo {title} {Structures and physical properties of \ce{R2TX3} compounds},\ }\href@noop {} {\bibfield  {journal} {\bibinfo  {journal} {Chinese Physics B}\ }\textbf {\bibinfo {volume} {22}},\ \bibinfo {pages} {056102} (\bibinfo {year} {2013})}\BibitemShut {NoStop}%
\bibitem [{\citenamefont {Mallik}\ \emph {et~al.}(1998)\citenamefont {Mallik}, \citenamefont {Sampathkumaran}, \citenamefont {Strecker}, \citenamefont {Wortmann}, \citenamefont {Paulose},\ and\ \citenamefont {Ueda}}]{Mallik1998}%
  \BibitemOpen
  \bibfield  {author} {\bibinfo {author} {\bibfnamefont {R.}~\bibnamefont {Mallik}}, \bibinfo {author} {\bibfnamefont {E.}~\bibnamefont {Sampathkumaran}}, \bibinfo {author} {\bibfnamefont {M.}~\bibnamefont {Strecker}}, \bibinfo {author} {\bibfnamefont {G.}~\bibnamefont {Wortmann}}, \bibinfo {author} {\bibfnamefont {P.}~\bibnamefont {Paulose}},\ and\ \bibinfo {author} {\bibfnamefont {Y.}~\bibnamefont {Ueda}},\ }\bibfield  {title} {\bibinfo {title} {Complex magnetism in a new alloy, \ce{Eu2PdSi3}, with two crystallographically inequivalent sites},\ }\href {https://doi.org/https://doi.org/10.1016/S0304-8853(98)00124-3} {\bibfield  {journal} {\bibinfo  {journal} {Journal of Magnetism and Magnetic Materials}\ }\textbf {\bibinfo {volume} {185}},\ \bibinfo {pages} {L135} (\bibinfo {year} {1998})}\BibitemShut {NoStop}%
\bibitem [{\citenamefont {Majumdar}\ \emph {et~al.}(1999)\citenamefont {Majumdar}, \citenamefont {Mallik}, \citenamefont {Sampathkumaran}, \citenamefont {Rupprecht},\ and\ \citenamefont {Wortmann}}]{Majumdar1999}%
  \BibitemOpen
  \bibfield  {author} {\bibinfo {author} {\bibfnamefont {S.}~\bibnamefont {Majumdar}}, \bibinfo {author} {\bibfnamefont {R.}~\bibnamefont {Mallik}}, \bibinfo {author} {\bibfnamefont {E.~V.}\ \bibnamefont {Sampathkumaran}}, \bibinfo {author} {\bibfnamefont {K.}~\bibnamefont {Rupprecht}},\ and\ \bibinfo {author} {\bibfnamefont {G.}~\bibnamefont {Wortmann}},\ }\bibfield  {title} {\bibinfo {title} {Magnetic behavior of ${\mathrm{eu}}_{2}{\mathrm{cusi}}_{3}:$ large negative magnetoresistance above the curie temperature},\ }\href {https://doi.org/10.1103/PhysRevB.60.6770} {\bibfield  {journal} {\bibinfo  {journal} {Phys. Rev. B}\ }\textbf {\bibinfo {volume} {60}},\ \bibinfo {pages} {6770} (\bibinfo {year} {1999})}\BibitemShut {NoStop}%
\bibitem [{\citenamefont {Cao}\ \emph {et~al.}(2010)\citenamefont {Cao}, \citenamefont {Klingeler}, \citenamefont {Vinzelberg}, \citenamefont {Leps}, \citenamefont {L\"oser}, \citenamefont {Behr}, \citenamefont {Muranyi}, \citenamefont {Kataev},\ and\ \citenamefont {B\"uchner}}]{cao2010}%
  \BibitemOpen
  \bibfield  {author} {\bibinfo {author} {\bibfnamefont {C.~D.}\ \bibnamefont {Cao}}, \bibinfo {author} {\bibfnamefont {R.}~\bibnamefont {Klingeler}}, \bibinfo {author} {\bibfnamefont {H.}~\bibnamefont {Vinzelberg}}, \bibinfo {author} {\bibfnamefont {N.}~\bibnamefont {Leps}}, \bibinfo {author} {\bibfnamefont {W.}~\bibnamefont {L\"oser}}, \bibinfo {author} {\bibfnamefont {G.}~\bibnamefont {Behr}}, \bibinfo {author} {\bibfnamefont {F.}~\bibnamefont {Muranyi}}, \bibinfo {author} {\bibfnamefont {V.}~\bibnamefont {Kataev}},\ and\ \bibinfo {author} {\bibfnamefont {B.}~\bibnamefont {B\"uchner}},\ }\bibfield  {title} {\bibinfo {title} {Magnetic anisotropy and ferromagnetic correlations above the curie temperature in \ce{Eu2CuSi3} single crystals},\ }\href {https://doi.org/10.1103/PhysRevB.82.134446} {\bibfield  {journal} {\bibinfo  {journal} {Phys. Rev. B}\ }\textbf {\bibinfo {volume} {82}},\ \bibinfo {pages} {134446} (\bibinfo {year} {2010})}\BibitemShut {NoStop}%
\bibitem [{\citenamefont {P\"{o}ttgen}\ and\ \citenamefont {Johrendt}(2000)}]{Pttgen2000}%
  \BibitemOpen
  \bibfield  {author} {\bibinfo {author} {\bibfnamefont {R.}~\bibnamefont {P\"{o}ttgen}}\ and\ \bibinfo {author} {\bibfnamefont {D.}~\bibnamefont {Johrendt}},\ }\bibfield  {title} {\bibinfo {title} {Equiatomic intermetallic europium compounds:{\hspace{0.167em}} syntheses, crystal chemistry, chemical bonding, and physical properties},\ }\href@noop {} {\bibfield  {journal} {\bibinfo  {journal} {Chemistry of Materials}\ }\textbf {\bibinfo {volume} {12}},\ \bibinfo {pages} {875} (\bibinfo {year} {2000})}\BibitemShut {NoStop}%
\bibitem [{\citenamefont {You}\ \emph {et~al.}(2007)\citenamefont {You}, \citenamefont {Grin},\ and\ \citenamefont {Miller}}]{you2007}%
  \BibitemOpen
  \bibfield  {author} {\bibinfo {author} {\bibfnamefont {T.-S.}\ \bibnamefont {You}}, \bibinfo {author} {\bibfnamefont {Y.}~\bibnamefont {Grin}},\ and\ \bibinfo {author} {\bibfnamefont {G.~J.}\ \bibnamefont {Miller}},\ }\bibfield  {title} {\bibinfo {title} {Planar versus puckered nets in the polar intermetallic series eugatt (tt = si, ge, sn)},\ }\href@noop {} {\bibfield  {journal} {\bibinfo  {journal} {Inorganic Chemistry}\ }\textbf {\bibinfo {volume} {46}},\ \bibinfo {pages} {8801} (\bibinfo {year} {2007})}\BibitemShut {NoStop}%
\bibitem [{\citenamefont {Zarechnyuk}\ and\ \citenamefont {Yanson}(1982)}]{Zarechnyuk1982}%
  \BibitemOpen
  \bibfield  {author} {\bibinfo {author} {\bibfnamefont {O.~S.}\ \bibnamefont {Zarechnyuk}}\ and\ \bibinfo {author} {\bibfnamefont {T.~I.}\ \bibnamefont {Yanson}},\ }\bibfield  {title} {\bibinfo {title} {Ternary phase diagram for the system {Al-Eu-Si} for the concentration range 0-33 at.\% {Eu}},\ }\href@noop {} {\bibfield  {journal} {\bibinfo  {journal} {Dopovidi Akademii Nauk Ukrains'koi RSR, Seriya B: Geologichni, Khimichni ta Biologichni Nauki}\ }\textbf {\bibinfo {volume} {4}},\ \bibinfo {pages} {31} (\bibinfo {year} {1982})}\BibitemShut {NoStop}%
\bibitem [{\citenamefont {Clark}\ and\ \citenamefont {Reid}(1995)}]{Clark1995}%
  \BibitemOpen
  \bibfield  {author} {\bibinfo {author} {\bibfnamefont {R.~C.}\ \bibnamefont {Clark}}\ and\ \bibinfo {author} {\bibfnamefont {J.~S.}\ \bibnamefont {Reid}},\ }\bibfield  {title} {\bibinfo {title} {{The analytical calculation of absorption in multifaceted crystals}},\ }\href {https://doi.org/10.1107/S0108767395007367} {\bibfield  {journal} {\bibinfo  {journal} {Acta Crystallogr. Sec. A}\ }\textbf {\bibinfo {volume} {51}},\ \bibinfo {pages} {887} (\bibinfo {year} {1995})}\BibitemShut {NoStop}%
\bibitem [{CrysAlisPro Software System; Rigaku Oxford Diffraction, vers. 1.171.42.75a; Rigaku Corporation, 2022()}]{Crysalispro}%
  \BibitemOpen
  CrysAlisPro Software System; Rigaku Oxford Diffraction, vers. 1.171.42.75a; Rigaku Corporation, 2022,\ \href@noop {} {}\BibitemShut {NoStop}%
\bibitem [{\citenamefont {Dolomanov}\ \emph {et~al.}(2009)\citenamefont {Dolomanov}, \citenamefont {Bourhis}, \citenamefont {Gildea}, \citenamefont {Howard},\ and\ \citenamefont {Puschmann}}]{Dolomanov2009}%
  \BibitemOpen
  \bibfield  {author} {\bibinfo {author} {\bibfnamefont {O.~V.}\ \bibnamefont {Dolomanov}}, \bibinfo {author} {\bibfnamefont {L.~J.}\ \bibnamefont {Bourhis}}, \bibinfo {author} {\bibfnamefont {R.~J.}\ \bibnamefont {Gildea}}, \bibinfo {author} {\bibfnamefont {J.~A.}\ \bibnamefont {Howard}},\ and\ \bibinfo {author} {\bibfnamefont {H.}~\bibnamefont {Puschmann}},\ }\bibfield  {title} {\bibinfo {title} {{OLEX2: A complete structure solution, refinement and analysis program}},\ }\href {https://doi.org/10.1107/S0021889808042726} {\bibfield  {journal} {\bibinfo  {journal} {J. Appl. Crystallogr.}\ }\textbf {\bibinfo {volume} {42}},\ \bibinfo {pages} {339} (\bibinfo {year} {2009})}\BibitemShut {NoStop}%
\bibitem [{\citenamefont {Sheldrick}(2015{\natexlab{a}})}]{Sheldrick2015}%
  \BibitemOpen
  \bibfield  {author} {\bibinfo {author} {\bibfnamefont {G.~M.}\ \bibnamefont {Sheldrick}},\ }\bibfield  {title} {\bibinfo {title} {{SHELXT - Integrated space-group and crystal-structure determination}},\ }\href {https://doi.org/10.1107/S2053273314026370} {\bibfield  {journal} {\bibinfo  {journal} {Acta Crystallogr. Sec. A}\ }\textbf {\bibinfo {volume} {71}},\ \bibinfo {pages} {3} (\bibinfo {year} {2015}{\natexlab{a}})}\BibitemShut {NoStop}%
\bibitem [{\citenamefont {Sheldrick}(2015{\natexlab{b}})}]{Sheldrick2015a}%
  \BibitemOpen
  \bibfield  {author} {\bibinfo {author} {\bibfnamefont {G.~M.}\ \bibnamefont {Sheldrick}},\ }\bibfield  {title} {\bibinfo {title} {{Crystal structure refinement with SHELXL}},\ }\href {https://doi.org/10.1107/S2053229614024218} {\bibfield  {journal} {\bibinfo  {journal} {Acta Crystallogr. Sec. C}\ }\textbf {\bibinfo {volume} {71}},\ \bibinfo {pages} {3} (\bibinfo {year} {2015}{\natexlab{b}})}\BibitemShut {NoStop}%
\bibitem [{\citenamefont {Spek}(2009)}]{Spek2009}%
  \BibitemOpen
  \bibfield  {author} {\bibinfo {author} {\bibfnamefont {A.~L.}\ \bibnamefont {Spek}},\ }\bibfield  {title} {\bibinfo {title} {{Structure validation in chemical crystallography}},\ }\href {https://doi.org/10.1107/S090744490804362X} {\bibfield  {journal} {\bibinfo  {journal} {Acta Crystallographica Section D}\ }\textbf {\bibinfo {volume} {65}},\ \bibinfo {pages} {148} (\bibinfo {year} {2009})}\BibitemShut {NoStop}%
\bibitem [{\citenamefont {Salamakha}\ \emph {et~al.}(2013)\citenamefont {Salamakha}, \citenamefont {Bauer}, \citenamefont {Hilscher}, \citenamefont {Michor}, \citenamefont {Sologub}, \citenamefont {Rogl},\ and\ \citenamefont {Giester}}]{Salamakha2013}%
  \BibitemOpen
  \bibfield  {author} {\bibinfo {author} {\bibfnamefont {L.}~\bibnamefont {Salamakha}}, \bibinfo {author} {\bibfnamefont {E.}~\bibnamefont {Bauer}}, \bibinfo {author} {\bibfnamefont {G.}~\bibnamefont {Hilscher}}, \bibinfo {author} {\bibfnamefont {H.}~\bibnamefont {Michor}}, \bibinfo {author} {\bibfnamefont {O.}~\bibnamefont {Sologub}}, \bibinfo {author} {\bibfnamefont {P.}~\bibnamefont {Rogl}},\ and\ \bibinfo {author} {\bibfnamefont {G.}~\bibnamefont {Giester}},\ }\bibfield  {title} {\bibinfo {title} {Structural and physical properties diversity of new \ce{CaCu5}-type related europium platinum borides},\ }\href@noop {} {\bibfield  {journal} {\bibinfo  {journal} {Inorganic Chemistry}\ }\textbf {\bibinfo {volume} {52}},\ \bibinfo {pages} {4185} (\bibinfo {year} {2013})}\BibitemShut {NoStop}%
\bibitem [{\citenamefont {Tian}\ \emph {et~al.}(2019)\citenamefont {Tian}, \citenamefont {Wang}, \citenamefont {Ji}, \citenamefont {Wang}, \citenamefont {Xia}, \citenamefont {Wang}, \citenamefont {Liu}, \citenamefont {Zhang},\ and\ \citenamefont {Cheng}}]{Tian2019}%
  \BibitemOpen
  \bibfield  {author} {\bibinfo {author} {\bibfnamefont {C.-K.}\ \bibnamefont {Tian}}, \bibinfo {author} {\bibfnamefont {C.}~\bibnamefont {Wang}}, \bibinfo {author} {\bibfnamefont {W.}~\bibnamefont {Ji}}, \bibinfo {author} {\bibfnamefont {J.-C.}\ \bibnamefont {Wang}}, \bibinfo {author} {\bibfnamefont {T.-L.}\ \bibnamefont {Xia}}, \bibinfo {author} {\bibfnamefont {L.}~\bibnamefont {Wang}}, \bibinfo {author} {\bibfnamefont {J.-J.}\ \bibnamefont {Liu}}, \bibinfo {author} {\bibfnamefont {H.-X.}\ \bibnamefont {Zhang}},\ and\ \bibinfo {author} {\bibfnamefont {P.}~\bibnamefont {Cheng}},\ }\bibfield  {title} {\bibinfo {title} {Domain wall pinning and hard magnetic phase in co-doped bulk single crystalline \ce{Fe3GeTe2}},\ }\href@noop {} {\bibfield  {journal} {\bibinfo  {journal} {Phys. Rev. B}\ }\textbf {\bibinfo {volume} {99}},\ \bibinfo {pages} {184428} (\bibinfo {year} {2019})}\BibitemShut {NoStop}%
\bibitem [{\citenamefont {G\'ornicka}\ \emph {et~al.}(2019)\citenamefont {G\'ornicka}, \citenamefont {Xie}, \citenamefont {Carnicom}, \citenamefont {Cava},\ and\ \citenamefont {Klimczuk}}]{Gornicka2019}%
  \BibitemOpen
  \bibfield  {author} {\bibinfo {author} {\bibfnamefont {K.}~\bibnamefont {G\'ornicka}}, \bibinfo {author} {\bibfnamefont {W.}~\bibnamefont {Xie}}, \bibinfo {author} {\bibfnamefont {E.~M.}\ \bibnamefont {Carnicom}}, \bibinfo {author} {\bibfnamefont {R.~J.}\ \bibnamefont {Cava}},\ and\ \bibinfo {author} {\bibfnamefont {T.}~\bibnamefont {Klimczuk}},\ }\bibfield  {title} {\bibinfo {title} {Synthesis and physical properties of the 10.6 {K} ferromagnet \ce {NdIr3}},\ }\href@noop {} {\bibfield  {journal} {\bibinfo  {journal} {Phys. Rev. B}\ }\textbf {\bibinfo {volume} {99}},\ \bibinfo {pages} {104430} (\bibinfo {year} {2019})}\BibitemShut {NoStop}%
\bibitem [{\citenamefont {Arrott}(1957)}]{Arrot1957}%
  \BibitemOpen
  \bibfield  {author} {\bibinfo {author} {\bibfnamefont {A.}~\bibnamefont {Arrott}},\ }\bibfield  {title} {\bibinfo {title} {Criterion for ferromagnetism from observations of magnetic isotherms},\ }\href@noop {} {\bibfield  {journal} {\bibinfo  {journal} {Phys. Rev.}\ }\textbf {\bibinfo {volume} {108}},\ \bibinfo {pages} {1394} (\bibinfo {year} {1957})}\BibitemShut {NoStop}%
\bibitem [{\citenamefont {Lef{\`{e}}vre}\ and\ \citenamefont {von Rohr}(2022)}]{Lefvre2022}%
  \BibitemOpen
  \bibfield  {author} {\bibinfo {author} {\bibfnamefont {R.}~\bibnamefont {Lef{\`{e}}vre}}\ and\ \bibinfo {author} {\bibfnamefont {F.~O.}\ \bibnamefont {von Rohr}},\ }\bibfield  {title} {\bibinfo {title} {A heavy fermion zn-deficient \ce{CaBe2Ge2}-type phase with rare {C}e-based ferromagnetism and large magnetoresistance},\ }\href@noop {} {\bibfield  {journal} {\bibinfo  {journal} {Chemistry of Materials}\ }\textbf {\bibinfo {volume} {34}},\ \bibinfo {pages} {2352} (\bibinfo {year} {2022})}\BibitemShut {NoStop}%
\bibitem [{\citenamefont {Pramanik}\ and\ \citenamefont {Banerjee}(2009)}]{Pramanik2009}%
  \BibitemOpen
  \bibfield  {author} {\bibinfo {author} {\bibfnamefont {A.~K.}\ \bibnamefont {Pramanik}}\ and\ \bibinfo {author} {\bibfnamefont {A.}~\bibnamefont {Banerjee}},\ }\bibfield  {title} {\bibinfo {title} {Critical behavior at paramagnetic to ferromagnetic phase transition in \ce{Pr_{0.5}Sr_{0.5}MnO3}: A bulk magnetization study},\ }\href@noop {} {\bibfield  {journal} {\bibinfo  {journal} {Phys. Rev. B}\ }\textbf {\bibinfo {volume} {79}},\ \bibinfo {pages} {214426} (\bibinfo {year} {2009})}\BibitemShut {NoStop}%
\bibitem [{\citenamefont {Fu}\ \emph {et~al.}(2020)\citenamefont {Fu}, \citenamefont {Qin}, \citenamefont {Sun}, \citenamefont {Hao}, \citenamefont {Zheng}, \citenamefont {Lohstroh}, \citenamefont {G\"{u}nther}, \citenamefont {Russina}, \citenamefont {Liu}, \citenamefont {Xiao}, \citenamefont {Jin},\ and\ \citenamefont {Chen}}]{fu2020}%
  \BibitemOpen
  \bibfield  {author} {\bibinfo {author} {\bibfnamefont {Z.}~\bibnamefont {Fu}}, \bibinfo {author} {\bibfnamefont {L.}~\bibnamefont {Qin}}, \bibinfo {author} {\bibfnamefont {K.}~\bibnamefont {Sun}}, \bibinfo {author} {\bibfnamefont {L.}~\bibnamefont {Hao}}, \bibinfo {author} {\bibfnamefont {Y.-Z.}\ \bibnamefont {Zheng}}, \bibinfo {author} {\bibfnamefont {W.}~\bibnamefont {Lohstroh}}, \bibinfo {author} {\bibfnamefont {G.}~\bibnamefont {G\"{u}nther}}, \bibinfo {author} {\bibfnamefont {M.}~\bibnamefont {Russina}}, \bibinfo {author} {\bibfnamefont {Y.}~\bibnamefont {Liu}}, \bibinfo {author} {\bibfnamefont {Y.}~\bibnamefont {Xiao}}, \bibinfo {author} {\bibfnamefont {W.}~\bibnamefont {Jin}},\ and\ \bibinfo {author} {\bibfnamefont {D.}~\bibnamefont {Chen}},\ }\bibfield  {title} {\bibinfo {title} {Low-temperature spin dynamics of ferromagnetic molecular ring \ce{Cr8Y8}},\ }\href@noop {} {\bibfield  {journal} {\bibinfo  {journal} {npj Quantum Mater.}\ }\textbf {\bibinfo {volume} {5}} (\bibinfo {year}
  {2020})}\BibitemShut {NoStop}%
\bibitem [{\citenamefont {Adhikari}\ \emph {et~al.}(2019)\citenamefont {Adhikari}, \citenamefont {Shen}, \citenamefont {Kunwar}, \citenamefont {Jeon}, \citenamefont {Maple}, \citenamefont {Dzero},\ and\ \citenamefont {Almasan}}]{Adhikari2019}%
  \BibitemOpen
  \bibfield  {author} {\bibinfo {author} {\bibfnamefont {R.~B.}\ \bibnamefont {Adhikari}}, \bibinfo {author} {\bibfnamefont {P.}~\bibnamefont {Shen}}, \bibinfo {author} {\bibfnamefont {D.~L.}\ \bibnamefont {Kunwar}}, \bibinfo {author} {\bibfnamefont {I.}~\bibnamefont {Jeon}}, \bibinfo {author} {\bibfnamefont {M.~B.}\ \bibnamefont {Maple}}, \bibinfo {author} {\bibfnamefont {M.}~\bibnamefont {Dzero}},\ and\ \bibinfo {author} {\bibfnamefont {C.~C.}\ \bibnamefont {Almasan}},\ }\bibfield  {title} {\bibinfo {title} {Magnetic field dependence of the schottky anomaly in filled skutterudites \ce{Pr_{1-x}Eu_{x}Pt4Ge12}},\ }\href@noop {} {\bibfield  {journal} {\bibinfo  {journal} {Phys. Rev. B}\ }\textbf {\bibinfo {volume} {100}},\ \bibinfo {pages} {174509} (\bibinfo {year} {2019})}\BibitemShut {NoStop}%
\bibitem [{\citenamefont {Kaczorowski}\ \emph {et~al.}(2012)\citenamefont {Kaczorowski}, \citenamefont {Belan},\ and\ \citenamefont {Gladyshevskii}}]{Kaczorowski2012}%
  \BibitemOpen
  \bibfield  {author} {\bibinfo {author} {\bibfnamefont {D.}~\bibnamefont {Kaczorowski}}, \bibinfo {author} {\bibfnamefont {B.}~\bibnamefont {Belan}},\ and\ \bibinfo {author} {\bibfnamefont {R.}~\bibnamefont {Gladyshevskii}},\ }\bibfield  {title} {\bibinfo {title} {Magnetic and electrical properties of \ce{EuPdGe3}},\ }\href@noop {} {\bibfield  {journal} {\bibinfo  {journal} {Solid State Communications}\ }\textbf {\bibinfo {volume} {152}},\ \bibinfo {pages} {839} (\bibinfo {year} {2012})}\BibitemShut {NoStop}%
\bibitem [{\citenamefont {Kumar}\ \emph {et~al.}(2011)\citenamefont {Kumar}, \citenamefont {Das}, \citenamefont {Kulkarni}, \citenamefont {Thamizhavel}, \citenamefont {Dhar},\ and\ \citenamefont {Bonville}}]{Kumar2011}%
  \BibitemOpen
  \bibfield  {author} {\bibinfo {author} {\bibfnamefont {N.}~\bibnamefont {Kumar}}, \bibinfo {author} {\bibfnamefont {P.~K.}\ \bibnamefont {Das}}, \bibinfo {author} {\bibfnamefont {R.}~\bibnamefont {Kulkarni}}, \bibinfo {author} {\bibfnamefont {A.}~\bibnamefont {Thamizhavel}}, \bibinfo {author} {\bibfnamefont {S.~K.}\ \bibnamefont {Dhar}},\ and\ \bibinfo {author} {\bibfnamefont {P.}~\bibnamefont {Bonville}},\ }\bibfield  {title} {\bibinfo {title} {Antiferromagnetic ordering in \ce{EuPtGe3}},\ }\href@noop {} {\bibfield  {journal} {\bibinfo  {journal} {Journal of Physics: Condensed Matter}\ }\textbf {\bibinfo {volume} {24}},\ \bibinfo {pages} {036005} (\bibinfo {year} {2011})}\BibitemShut {NoStop}%
\bibitem [{\citenamefont {Tari}(2003)}]{Tari2003}%
  \BibitemOpen
  \bibfield  {author} {\bibinfo {author} {\bibfnamefont {A.}~\bibnamefont {Tari}},\ }\href {https://doi.org/10.1142/p254} {\emph {\bibinfo {title} {The Specific Heat of Matter at Low Temperatures}}}\ (\bibinfo  {publisher} {Imperial Collage Press},\ \bibinfo {year} {2003})\BibitemShut {NoStop}%
\bibitem [{\citenamefont {Blundell}(2001)}]{Blundell}%
  \BibitemOpen
  \bibfield  {author} {\bibinfo {author} {\bibfnamefont {S.}~\bibnamefont {Blundell}},\ }\href@noop {} {\emph {\bibinfo {title} {Magnetism in Condensed Matter}}}\ (\bibinfo  {publisher} {Oxford Master Series in Physics},\ \bibinfo {year} {2001})\BibitemShut {NoStop}%
\bibitem [{\citenamefont {Mudiyanselage}\ \emph {et~al.}(2023)\citenamefont {Mudiyanselage}, \citenamefont {Klimczuk},\ and\ \citenamefont {Xie}}]{klimczuk2023}%
  \BibitemOpen
  \bibfield  {author} {\bibinfo {author} {\bibfnamefont {R.~S.~D.}\ \bibnamefont {Mudiyanselage}}, \bibinfo {author} {\bibfnamefont {T.}~\bibnamefont {Klimczuk}},\ and\ \bibinfo {author} {\bibfnamefont {W.}~\bibnamefont {Xie}},\ }\bibfield  {title} {\bibinfo {title} {Critical charge transfer pairs in intermetallic superconductors},\ }in\ \href@noop {} {\emph {\bibinfo {booktitle} {Comprehensive Inorganic Chemistry {III}}}}\ (\bibinfo  {publisher} {Elsevier},\ \bibinfo {year} {2023})\ pp.\ \bibinfo {pages} {217--235}\BibitemShut {NoStop}%
\bibitem [{\citenamefont {Matthias}\ \emph {et~al.}(1958)\citenamefont {Matthias}, \citenamefont {Suhl},\ and\ \citenamefont {Corenzwit}}]{Matthias1958}%
  \BibitemOpen
  \bibfield  {author} {\bibinfo {author} {\bibfnamefont {B.~T.}\ \bibnamefont {Matthias}}, \bibinfo {author} {\bibfnamefont {H.}~\bibnamefont {Suhl}},\ and\ \bibinfo {author} {\bibfnamefont {E.}~\bibnamefont {Corenzwit}},\ }\bibfield  {title} {\bibinfo {title} {Spin exchange in superconductors},\ }\href {https://doi.org/10.1103/PhysRevLett.1.92} {\bibfield  {journal} {\bibinfo  {journal} {Phys. Rev. Lett.}\ }\textbf {\bibinfo {volume} {1}},\ \bibinfo {pages} {92} (\bibinfo {year} {1958})}\BibitemShut {NoStop}%
\end{thebibliography}%

\end{document}